\documentclass[%
reprint,
showpacs,
 amsmath,amssymb,
aps,
]{revtex4-1}

\usepackage{graphicx}
\usepackage{dcolumn}
\usepackage{bm}

\begin{document}

\preprint{APS/123-QED}

\title{Coupled-wave model for square-lattice photonic-crystal lasers with TE polarization \---- a general approach}

\author{Yong Liang, Chao Peng,$^{*}$ Kyosuke Sakai, Seita Iwahashi, and Susumu Noda$^{\dag}$}

  \affiliation{%
 Department of Electronic Science and Engineering, Kyoto University, Kyoto-Daigaku-Katsura, Nishikyo-ku, Kyoto 615-8510, Japan\\
 E-mail: $^{*}$pengchao@qoe.kuee.kyoto-u.ac.jp, $^{\dag}$snoda@kuee.kyoto-u.ac.jp
}%

\date{\today}

\begin{abstract}
A general coupled-wave model is presented for square-lattice photonic crystal (PC) lasers with transverse-electric polarization. This model incorporates the high-order coupling effects that are important for two-dimensional PC laser cavities and gives a general and rigorous coupled-wave formulation for the full three-dimensional structures of typical laser devices. Numerical examples based on our model are presented for PC structures with different air-hole shapes. The accuracy of the results obtained is verified using three-dimensional finite-difference time-domain simulations.

\end{abstract}

\maketitle


\section{\label{sec:level1}Introduction}
Two-dimensional (2D) photonic crystal surface emitting lasers (PC-SELs) are becoming increasingly important due to their enhanced functionality and improved performance compared to conventional semiconductor lasers \cite{imada1}-\cite{kurosaka1}. By utilizing the band edge of the photonic band structure, single longitudinal and transverse mode oscillation in two dimensions has been achieved with a large lasing area, enabling high-power, single-mode operation \cite{imada1,ohnishi}. The output beam of such devices is emitted in the direction normal to the 2D PC plane due to first-order Bragg diffraction effects. Importantly, this surface-emitted beam emanating from a large lasing area has a small beam divergence angle \cite{ohnishi,miyai1}. Furthermore, both the polarization and pattern of the output beam can be controlled by appropriate design of the PC geometry \cite{noda-science,miyai1}. Recent developments of 2D PC lasers have allowed the lasing wavelength to be extended from the near-infrared regime to the mid-infrared \cite{kim}, terahertz \cite{sirigu,chassagneux}, and blue-violet regimes \cite{matsubara,lu}. In addition, we have recently demonstrated the operation of a PC-SEL with entirely new functionality: on-chip dynamical control of the emitted beam direction, achieved by using a composite PC structure \cite{kurosaka1}.

Despite the experimental advances that have recently been made in the field of 2D PC-SELs, theoretical studies on these types of lasers have thus far been limited. An important but unresolved issue concerns the mechanisms by which the PC structure determines the output characteristics of the device, thereby limiting progress in optimizing the structural design of devices. Computer simulations based on the 2D plane-wave expansion method (PWEM) \cite{imada2,maradudin} or the finite-difference time-domain (FDTD) method \cite{fan,notomi,yokoyama}, can provide valuable information about the lasing properties of the PC laser cavity. However, there are some inherent limitations to these computational approaches. The 2D PWEM is only applicable to infinite structures, and the FDTD method requires substantial computational resources in order to model finite structures with realistically large areas. Moreover, neither simulation approach provides deep analytical insight into improving the design of devices. A group of alternative analytical methods \cite{vurgaftman,toda} have been developed, based on the concept of one-dimensional (1D) coupled-wave theory (CWT) \cite{kogelnik}. However, these methods in their initial formulations consider only four basic waves in the coupled-wave model and disregard 2D optical coupling effects, which are important in 2D square-lattice PCs. The consequent shortcomings do not allow accurate explanations of experimental results to be obtained \cite{ohnishi}. Sakai et al. later derived a 2D CWT using an eight-wave model \cite{sakai1,sakai2} that incorporates both conventional 1D coupling and the more recently pertinent 2D coupling,   allowing 2D coherent lasing action to be explained \cite{ohnishi}. This more detailed formulation of the lasing process underscores the importance of accurately modeling complicated 2D optical coupling effects in 2D PC-SELs.

Nevertheless, two fundamental limitations remain in the 2D CWT approach \cite{sakai1,sakai2}.
First, light propagating within the PC is in principle a Bloch wave described by an infinite number of terms in a Fourier series expansion, and thus it is crucial to include as many wave orders as possible to model the PC cavity accurately. The eight-wave model in the CWT mentioned above is applicable only to PCs in which the air holes have simple symmetric shapes, such as circular holes. However, it is becoming increasingly clear that the use of air holes with \emph{asymmetric} shapes, such as equilateral triangles and right-angled isosceles triangles \cite{kunishi,sakaguchi}, is often beneficial; these shapes  are potentially important for improving the output power and slope efficiency of 2D PC-SELs. Therefore, an extension of the coupled-wave model to the analysis of more complicated PC geometries requires the inclusion of more higher-order terms.
The second limitation of the existing 2D CWT is that the approach is confined to 2D models in which the structure is assumed to be uniform in the vertical direction. However, realistic PC-SELs require three-dimensional (3D) analysis because the waveguide structure breaks the structural uniformity in the vertical direction. Although effective dielectric constants of the PC materials \cite{imada2,sakai0} have previously been used to compensate for relatively small optical confinement in the vertical direction, this approach is inadequate to model the 3D structure in cases where the transverse field profile of the individual waves must be treated very carefully, as will be shown below. Thus, it is necessary to derive a more rigorous formulation by considering a 3D system instead of a 2D system.

In this paper, we develop a more general coupled-wave model for square-lattice PC-SELs with transverse-electric (TE) polarization in order to overcome the limitations discussed above.
Our coupled-wave model incorporates a large number of high-order wavevectors in order to capture all the important coupling effects. We present a rigorous coupled-wave formulation for a 3D structure by extending the general coupled-wave approach developed by Streifer \textit{et al.} for 1D distributed feedback (DFB) lasers \cite{streifer}.
This formulation not only models the coupling effects in 3D systems more accurately, but can also be generally applied to the analysis of PC structures with air holes of arbitrary shape.
We present numerical examples in which the mode frequency and radiation constant of PC structures with several different air-hole shapes are calculated and compared with the results of 3D-FDTD simulations in order to confirm the validity and accuracy of the extended CWT model.

This paper is organized as follows. Section \textbf{II} describes the coupled-wave model, and derivations of the coupled-wave equations are given. Section \textbf{III} presents our numerical results, which we then discuss. A summary of our findings is given in Section \textbf{IV}.

\section{Formulation of coupled-wave model}\label{sec:level2l}

\begin{figure}[htbp]
\centerline{
\includegraphics[width=8cm]{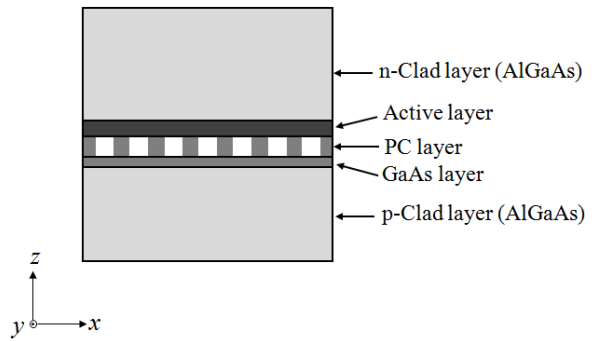}}
 \caption{Schematic cross-sectional view of a PC-SEL. This multilayer waveguide structure represents an approximation of a realistic laser device.}\label{fig.waveguide}
\end{figure}

\begin{figure}[htbp]
  \includegraphics[width=6cm]{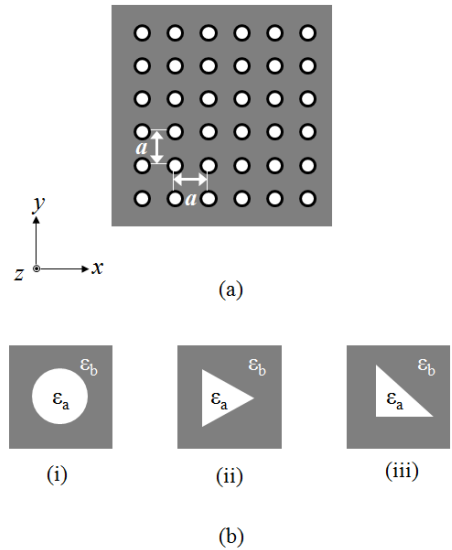}\\
  \caption{(a) Schematic diagram of a square-lattice photonic crystal. (b) Examples of air-hole designs: (i) circular, (ii) equilateral triangular, (iii) right-angled isosceles triangular.}
  \label{fig.sq-lattice}
\end{figure}

A schematic cross-section of the PC-SEL device considered here is shown in Fig. \ref{fig.waveguide}, which can be approximated by a multilayer waveguide. The PC layer is embedded in the waveguide structure, which is assumed to support only a single waveguide mode. The structural parameters of each layer are summarized in Table \ref{Tab.1}. The average dielectric constant of the PC layer is given by $\varepsilon_{av}=f\cdot\varepsilon_a+(1-f)\cdot\varepsilon_b$, where $\varepsilon_a$ is the dielectric constant of air, $\varepsilon_b$  is the dielectric constant of the background dielectric material (GaAs), $f$ is the filling factor (FF) given by  $f=S_{air-hole}/a^2$ (i.e., the fraction of the area of a unit cell occupied by air holes), and $a$ is the lattice constant. The PC layer consists of a square lattice with air holes perpendicular to the $xy$ plane, as shown in Fig. \ref{fig.sq-lattice}(a). In this paper, the shape of the air-holes is not restricted to circular but can be arbitrary. Figure \ref{fig.sq-lattice}(b) depicts examples of air-hole shapes that are considered later in this paper: (i) circles, (ii) equilateral triangles, and (iii) right-angled isosceles triangles.

\begin{table}[htbp]
\centering \tabcolsep 3mm \caption{Waveguide structural parameters}\label{Tab.1}
\begin{tabular}{c c c}
\hline
 Layer &  Thickness ($a$) & Dielectric constant \\
  \hline\hline
  n-clad(AlGaAs) & $\infty$& 11.0224 \\
  Active & 0.3 & 12.8603 \\
  PC & 0.4 & $\varepsilon_{av}$ \\
  GaAs & 0.2 & 12.7449 \\
  p-clad(AlGaAs) & $\infty$ & 11.0224 \\
  \hline
\end{tabular}
\end{table}

\begin{figure}[htbp]
  \includegraphics[width=7cm]{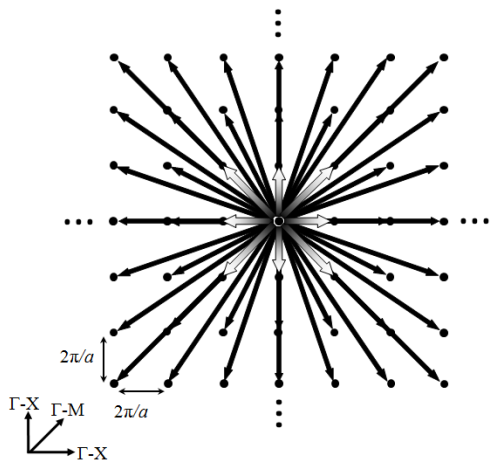}\\
  \caption{Bloch wave state represented by wavevectors (arrows) in reciprocal space. A large number of high-order wavevectors are included, in addition to the eight wavevectors around the center (shaded arrows), which were considered in the previous CWT model.}
  \label{fig.CWTmodel}
\end{figure}

Light propagating inside a PC must obey Bloch's theorem, which states that the amplitude of the light must conform to the imposed periodicity \cite{engelen}. The Bloch wave state $\psi$ can be expressed as
\begin{eqnarray}
\label{Eq0.1}
\psi_k(\bm r) = u_k(x,y,z)e^{-i\bm k \cdot \bm r}= \sum_{\bm G_{m,n}} a_{\bm G_{m,n}}(z)e^{-i (\bm k+\bm G_{m,n})\cdot \bm r } \nonumber \\
\end{eqnarray}
where the periodicity implies that $u_k(x,y,z)=u_k(x+a,y,z)=u_k(x,y+a,z)$. Here, $\bm k$ is a wavevector in the first Brillouin zone; $a_{\bm G_{m,n}}$ is the field amplitude of a given reciprocal lattice vector,  defined as $\bm G_{m,n}=(m\beta_0, n\beta_0)$ where $m$ and $n$ are arbitrary integers, and $\beta_0=2\pi/a$. At the second-order $\Gamma$-point $\bm k$ becomes zero \cite{sakai1}, thus Eq. (\ref{Eq0.1}) can be rewritten as a Fourier series of plane waves:
\begin{eqnarray}
\label{Eq0.2}
\psi(\bm r) = \sum_{m,n}a_{m,n}(z)e^{-i (m\beta_0 x+n\beta_0 y) },
\end{eqnarray}
where $a_{m,n}$ represents the field amplitude of a wave with wavevector $(m\beta_0, n\beta_0)$. It is implied by Eq. (\ref{Eq0.2}) that the Bloch wave excited in the vicinity of the $\Gamma$-point is composed of multiple wavevectors, including the wavevectors indicated by arrows in the reciprocal space diagram in Fig. \ref{fig.CWTmodel}, as well as wavevectors outside the plotted range. The shaded arrows close to the center of the plot ($m^2+n^2\leq 2$) indicate the eight waves considered in the existing CWT approach. In order to make our analysis generally applicable to air-hole geometries that require a large number of Fourier coefficients to model accurately, we will extend the existing coupled-wave model by including not only these eight wavevectors but also many high-order wavevectors.

\begin{figure}[htbp]
  \includegraphics[width=8cm]{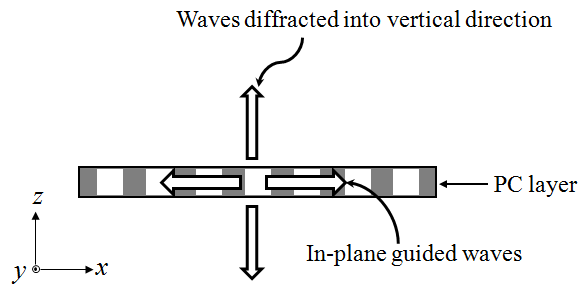}\\
  \caption{Schematic diagram showing the in-plane guided waves and waves diffracted vertical to the PC plane.}\label{fig.radiative}
\end{figure}

For TE polarization, the Bloch wave state $\psi$ can be described by either the magnetic-field component ($H_z$) or the electric-field components ($E_x, Ey$). At the second-order $\Gamma$-point, not only the in-plane guided waves but also waves that are diffracted into the vertical ($z$) direction exist, as shown in Fig. \ref{fig.radiative} \cite{ohnishi}. It is important to include these diffracted waves in the coupled-wave model because they determine the output (loss) of the PC laser cavity. We choose to start our formulation from MaxwellÕs equations using the electric-field components, because scalar wave equations for the magnetic field ($H_z$) \cite{sakai1} cannot include waves diffracted in the vertical direction (we note that there is no $H_z$ component for these waves).

In a 3D waveguide structure, as shown in Fig. \ref{fig.waveguide}, the electric field can be expressed as $\bm E(\bm r,t) = \bm E(\bm r)e^{iwt}$. By eliminating the magnetic field from Maxwell's equations, we obtain
\begin{eqnarray}
\label{Eq1.1} \nabla \times \nabla \times \bm{E}(\bm r) =
k_0^2\tilde{n}^2(\bm r)\bm{E}(\bm r),
\end{eqnarray}
where $k_0(=\omega/c)$ is the free-space wavenumber, $\omega$ is the angular frequency, $c$ is the velocity of light in free space and $\tilde{n}$ is the refractive index (a complex number) that satisfies \cite{streifer}
\begin{eqnarray}
\label{Eq1.2}
k_0^2\tilde{n}^2(\bm r)=k_0^2n^2(\bm r) + 2ik_0n(\bm r)\tilde{\alpha}(\bm r)-\tilde{\alpha}^2(\bm r),
\end{eqnarray}
where $n(\bm r)$ is the real part of $\tilde{n}(\bm r)$ and $\tilde{\alpha}(z)$ represents the gain ($\tilde{\alpha}>0$) or loss ($\tilde{\alpha}<0$) in each region. In the following derivation, we neglect the third term $\tilde{\alpha}^2(z)$ because $|\tilde{\alpha}|<<k_0n_0$ \cite{streifer}. For TE polarization $\bm E(\bm r)= (E_x(\bm r), E_y(\bm r), 0)$, and we obtain the following three scalar equations from Eq. (\ref{Eq1.1}):
\begin{eqnarray}
\label{Eq2.1}
&&[\frac{\partial^2 }{\partial z^2} + \frac{\partial^2 }{\partial
y^2} + k_0^2\tilde{n}^2]E_x - \frac{\partial ^2}{\partial x
\partial y}E_y = 0,
\\\label{Eq2.2} &&[\frac{\partial^2 }{\partial z^2} +
\frac{\partial^2 }{\partial x^2} + k_0^2\tilde{n}^2]E_y -
\frac{\partial ^2}{\partial x \partial y}E_x = 0,
\\\label{Eq2.3} && \frac{\partial }{\partial z} [\frac{\partial E_x }{\partial x} +
\frac{\partial E_y }{\partial y}] = 0.
\end{eqnarray}

In order to solve these equations, we expand $E_x(\bm r), E_y(\bm r)$ and $n^2(\bm r)$ according to
\begin{eqnarray}
\label{Eq3.1} &&E_x(\bm r) = \sum_{m,n} E_{x,m,n}(z)e^{-i\beta_m x - i\beta_n y},\\
\label{Eq3.2} &&E_y(\bm r) = \sum_{m,n} E_{y,m,n}(z)e^{-i\beta_m x - i\beta_n y},\\
\label{Eq3.3} &&n^2(\bm r) = n_0^2(z) + \sum_{\substack{m \neq 0,\\ n \neq 0}}
\xi_{m,n}(z)e^{-i\beta_m x - i\beta_n y}.
\end{eqnarray}
Here, $\beta_m = m\beta_0$, $\beta_n = n\beta_0$, $ \beta_0=2\pi/a$ ($m, n$ are arbitrary integers), $n_0^2(z)=\varepsilon_0(z)$, $\varepsilon_0(z)$ is the average dielectric constant of the material at position $z$, and $\xi_{m,n}(z)$ is the high-order Fourier coefficient term. We note that $\xi_{m,n}(z)$ is zero outside the PC region. Inside the PC region, $n_0^2(z)$ and $\xi_{m,n}(z)$ can be expressed as
\begin{eqnarray}
\label{Eq3.4} n_0^2(z) &&=\varepsilon_{av}=f\cdot\varepsilon_a+(1-f)\cdot\varepsilon_b, \\
\label{Eq3.5} \xi_{m,n}(z) &&= \frac{1}{a^2}\iint_{-a/2}^{a/2}n^2(x, y)e^{i\beta_m x + i\beta_n y} dxdy.
\end{eqnarray}
In order to simplify the formulation, we have assumed that air holes within the PC region have perfectly vertical sidewalls (a formulation for air holes with tilted sidewalls will be given elsewhere) such that $n_0^2$ and $\xi_{m,n}$ are independent of $z$ within this region.

By substituting Eqs. (\ref{Eq1.2}) and (\ref{Eq3.1})-(\ref{Eq3.3}) into Eqs. (\ref{Eq2.1})-(\ref{Eq2.3}), and collecting all terms that are multiplied by the factor $e^{-i\beta_mx-i\beta_ny}$, we obtain
\begin{subequations}
\begin{eqnarray}
\label{Eq4.1}
&&[\frac{\partial^2}{\partial z^2}+k_0^2n_0^2+2i\tilde{\alpha}k_0n_0(z)-n^2\beta_0^2]E_{x,m,n}+mn\beta_0^2E_{y,m,n}\nonumber \\
&&= -k_0^2\sum_{\substack{m'\neq m,\\n' \neq n}}\xi_{\substack{m-m',\\n-n'}}E_{x,m',n'}, \\
\label{Eq4.2}
&&[\frac{\partial^2}{\partial z^2}+k_0^2n_0^2+2i\tilde{\alpha}k_0n_0(z)-m^2\beta_0^2]E_{y,m,n}+mn\beta_0^2E_{x,m,n}\nonumber \\
&&= -k_0^2\sum_{\substack{m'\neq m,\\n' \neq n}}\xi_{\substack{m-m',\\n-n'}}E_{y,m',n'}, \\
\label{Eq4.3}
&&\frac{\partial}{\partial z}[mE_{x,mn}+nE_{y,mn}]=0.
\end{eqnarray}
\end{subequations}
In this work, the derivative terms of $E_x$ and $E_y$ with respect to $x$ and $y$ have been eliminated because we assume an infinite periodic PC structure (the corresponding formulation for a finite periodic PC structure will be given elsewhere).

The wavevectors can be classified into three groups according to their in-plane wavenumber, $\sqrt{m^2+n^2}\beta_0$.
\begin{itemize}
  \item Basic waves: $\sqrt{m^2+n^2}=1$,
  \item High-order waves: $\sqrt{m^2+n^2}>1$,
  \item Radiative waves: $m=0, n=0$.
\end{itemize}
In the resonant case at the second-order $\Gamma$-point \cite{sakai1,sakai2}, we assume that the basic waves can be expressed as
\begin{subequations}
\begin{eqnarray}
\label{Eq5.1}
E_{x,1,0} &=& 0,  \quad E_{y,1,0} = R_x\Theta_0(z),  \\
\label{Eq5.2}
E_{x,-1,0} &=& 0, \quad E_{y,-1,0} = S_x\Theta_0(z), \\
\label{Eq5.3}
E_{x,0,1} &=&  R_y\Theta_0(z), \quad E_{y,0,1}=0,   \\
\label{Eq5.4}
E_{x,0,-1} &=&  S_y\Theta_0(z),\quad  E_{y,0,-1}=0.
\end{eqnarray}
\end{subequations}
Here, $R_x$ and $S_x$ represent the amplitudes of basic waves propagating in the $+x$ and $-x$ directions, respectively, and likewise $R_y$ and $S_y$ represent the amplitudes of waves propagating in the $+y$ and $-y$ directions, respectively. These four basic waves are assumed to have identical field profiles in the $z$-direction, denoted by $\Theta_0(z)$, which is the same as the field profile of the fundamental waveguide mode for a waveguide with no periodic structure \cite{streifer, kazarinov}. We express the wave equation for the fundamental waveguide mode in terms of $\Theta_0(z)$ as
\begin{equation}
\label{Eq6} \frac{\partial^2 \Theta_0}{\partial z^2} + [k_0^2n_0^2(z) -\beta^2]\Theta_0=0,
\end{equation}
where $\beta$ is the propagation constant. The solutions for $\beta$ and $\Theta_0(z)$ in Eq. (\ref{Eq6}) can be obtained by employing the transfer matrix method (TMM) \cite{bergmann}.

In order to obtain the equations satisfied by the basic waves, Eqs. (\ref{Eq5.1})-(\ref{Eq5.4}) are substituted into Eqs. (\ref{Eq4.1})-(\ref{Eq4.3}) for $(m, n)=\{(1,0), (-1,0), (0,1), (0,-1)\}$.
Without any loss of generality, we focus here on the case where $(m,n)=(1,0)$. We then only need to consider Eqs. (\ref{Eq5.1}) and (\ref{Eq4.2}). Substitution of Eq. (\ref{Eq5.1}) into Eq. (\ref{Eq4.2}) gives
\begin{eqnarray}
\label{Eq7}
[\frac{\partial^2 \Theta_0}{\partial z^2} + (k_0^2n_0^2 + 2i\tilde{\alpha}k_0n_0(z)- \beta_0^2)\Theta_0]R_x \nonumber \\
=-k_0^2\sum_{\substack{m'\neq 1,\\n' \neq 0}}\xi_{\substack{1-m',\\-n'}}E_{y,m',n'}.
\end{eqnarray}
Next, Eq. (\ref{Eq6}) is substituted into Eq. (\ref{Eq7}) to yield
\begin{eqnarray}
\label{Eq8}
(\beta^2-\beta_0^2)R_x\Theta_0 &+& 2i\tilde{\alpha}k_0n_0(z)R_x\Theta_0\nonumber \\
&=&  -k_0^2\sum_{\substack{m'\neq 1,\\n' \neq 0}}\xi_{\substack{1-m',\\-n'}}E_{y,m',n'}.
\end{eqnarray}
Specifically, we express the radiative waves, i.e., the field amplitudes of the waves with $(m,n)=(0,0)$ as
\begin{eqnarray}
\label{Eq9}
E_{x,0,0}=\Delta E_x(z), E_{y,0,0}=\Delta E_y(z).
\end{eqnarray}

Finally, we can obtain the coupled-wave equation for $(m,n)=(1,0)$ bymultiplying Eq. (\ref{Eq8}) by $\Theta_0^*(z)$ on both sides and integrating over $(-\infty, \infty)$  along the $z$ direction. Three more coupled-wave equations for $(m,n)=\{(-1,0),(0,1),(0,-1)\}$ can be derived in analogous fashion. We write the four coupled-wave equations in the following form:
\begin{subequations}
\begin{eqnarray}
\label{Eq10.1}
(\delta+&&i\alpha)R_x =\kappa_{2,0}S_x\nonumber\\
&&-\frac{k_0^2}{2\beta_0P}\xi_{1,0}\int_{PC}\Delta E_y(z)\Theta_0^*(z)dz\\
&&-\frac{k_0^2}{2\beta_0P}\sum_{\sqrt{m^2+n^2}>1}\xi_{\substack{1-m,\\-n}}\int_{PC}E_{y,m,n}(z)\Theta_0^*(z)dz\nonumber,
\end{eqnarray}
\begin{eqnarray}
\label{Eq10.2}
(\delta+&&i\alpha)S_x =
\kappa_{-2,0}R_x\nonumber\\
&&-\frac{k_0^2}{2\beta_0P}\xi_{-1,0}\int_{PC}\Delta E_y(z)\Theta_0^*(z)dz\\
&&-\frac{k_0^2}{2\beta_0P}\sum_{\sqrt{m^2+n^2}>1}\xi_{\substack{-1-m,\\-n}}\int_{PC}E_{y,m,n}(z)\Theta_0^*(z)dz, \nonumber
\end{eqnarray}
\begin{eqnarray}
\label{Eq10.3}
(\delta+&&i\alpha)R_y =
\kappa_{0,2}S_y\nonumber\\
&&-\frac{k_0^2}{2\beta_0P}\xi_{0,1}\int_{PC}\Delta E_x(z)\Theta_0^*(z)dz\\
&&-\frac{k_0^2}{2\beta_0P}\sum_{\sqrt{m^2+n^2}>1}\xi_{\substack{-m,\\1-n}}\int_{PC}E_{x,m,n}(z)\Theta_0^*(z)dz, \nonumber
\end{eqnarray}
\begin{eqnarray}
\label{Eq10.4}
(\delta+&&i\alpha)S_y =
\kappa_{0,-2}R_y\nonumber\\
&&-\frac{k_0^2}{2\beta_0P}\xi_{0,-1}\int_{PC}\Delta E_x(z)\Theta_0^*(z)dz\\
&&-\frac{k_0^2}{2\beta_0P}\sum_{\sqrt{m^2+n^2}>1}\xi_{\substack{-m,\\-1-n}}\int_{PC}E_{x,m,n}(z)\Theta_0^*(z)dz,\nonumber
\end{eqnarray}
\end{subequations}
Here,
\begin{equation}
\label{Eq11}
\delta=\beta-\beta_0=n_{eff}(\omega-\omega_0)/c
\end{equation}
is the deviation from the Bragg condition, $\omega_0$ is the Bragg frequency, $n_{eff}$ is the effective refractive index of the PC layer, and $\alpha$ is the mode gain/loss given by
\begin{equation}
\alpha=\frac{k_0}{\beta_0P}\int_{-\infty}^\infty n_0(z)\tilde{\alpha}(z)|\Theta_0(z)|^2dz,
\end{equation}
where $P$ is a normalization factor given by
\begin{equation}
P=\int_{-\infty}^\infty |\Theta_0(z)|^2dz.
\end{equation}
The parameters $\kappa_{\pm2,0}, \kappa_{0,\pm2}$ are the conventional 1D (forward-backward) coupling coefficients given by
\begin{subequations}
\begin{eqnarray}
\label{Eq12.1}
\kappa_{\pm2,0}=-\frac{k_0^2}{2\beta_0P}\xi_{\pm2,0}\int_{PC}|\Theta_0(z)|^2dz,\\
\label{Eq12.2}
\kappa_{0,\pm2}=-\frac{k_0^2}{2\beta_0P}\xi_{0,\pm2}\int_{PC}|\Theta_0(z)|^2dz.
\end{eqnarray}
\end{subequations}
The integrals in Eqs. (\ref{Eq12.1})-(\ref{Eq12.2}), as well as those in Eqs. (\ref{Eq10.1})-(\ref{Eq10.4}), extend only over the PC region because $\xi_{mn}=0$ outside that range.

As the fields of the radiative waves ($\Delta E_x(z)$, $\Delta E_y(z)$) and the high-order waves ($E_{x,m,n}(z)$, $E_{y,m,n}(z)$)  are unknown, we cannot yet evaluate the right-hand sides of Eqs. (\ref{Eq10.1})-(\ref{Eq10.4}).
In order to determine these fields, Eqs. (\ref{Eq4.1})-(\ref{Eq4.3}) must be solved for these waves.
We will assume, following Ref. \cite{streifer}, that: (1) Only basic waves are important in generating radiative waves and high-order waves; (2) $\tilde{\alpha}$ is small and thus may be neglected.
These assumptions allow us to modify Eqs. (\ref{Eq4.1})-(\ref{Eq4.3}) as follows:
\begin{subequations}
\begin{eqnarray}
\label{Eq13.1}
&&[\frac{\partial^2}{\partial z^2}+k_0^2n_0^2-n^2\beta_0^2]E_{x,m,n}+mn\beta_0^2E_{y,m,n}\nonumber \\
&&= -k_0^2\sum_{\substack{m'\neq m,\\n' \neq n}}\xi_{\substack{m-m',\\n-n'}}E_{x,m',n'}, \\
\label{Eq13.2}
&&[\frac{\partial^2}{\partial z^2}+k_0^2n_0^2-m^2\beta_0^2]E_{y,m,n}+mn\beta_0^2E_{x,m,n}\nonumber \\
&&= -k_0^2\sum_{\substack{m'\neq m,\\n' \neq n}}\xi_{\substack{m-m',\\n-n'}}E_{y,m',n'}, \\
\label{Eq13.3}
&&\frac{\partial}{\partial z}[mE_{x,mn}+nE_{y,mn}]=0.
\end{eqnarray}
\end{subequations}
Here, $(m',n')=\{(1,0),(-1,0),(0,1),(0,-1)\}$ and $(m,n)$ is limited to the cases of radiative waves and high-order waves.
Below, we describe how these equations can be used to obtain solutions for radiative waves and high-order waves.

First, we consider the radiative waves $\Delta E_x(z)$ and $\Delta E_{y}(z)$ for $(m, n)=(0, 0)$. In this case, Eqs. (\ref{Eq13.1})-(\ref{Eq13.3}) are reduced to the following two expressions:
\begin{subequations}
\begin{eqnarray}
\label{Eq14.1}
[\frac{\partial^2 }{\partial z^2} + k_0^2n_0^2(z)]&&\Delta E_x(z)
=-k_0^2\sum_{m', n' \neq 0} \xi_{-m',-n'}E_{x,m',n'}\nonumber \\
&&\simeq-k_0^2(\xi_{0,-1}R_y+\xi_{0,1}S_y)\Theta_0(z),
\\
\label{Eq14.2}
[\frac{\partial^2 }{\partial z^2} + k_0^2n_0^2(z)]&&\Delta E_y(z)
=-k_0^2\sum_{m', n' \neq 0} \xi_{-m',-n'}E_{y,m',n'}\nonumber \\
&&\simeq-k_0^2(\xi_{-1,0}R_x+\xi_{1,0}S_x)\Theta_0(z).
\end{eqnarray}
\end{subequations}

These equations can be solved by employing the Green function approach \cite{kazarinov}, where the Green function $G(z,z')$ satisfies
\begin{subequations}
\begin{eqnarray}
\label{Eq15.1}
[\frac{\partial^2}{\partial z^2} + k_0^2n_0^2] G(z,z') &=& -\delta(z,z'),\\
\label{Eq15.2}
G(z,z') &\simeq& -i \cdot \frac{e^{-i\beta_z|z-z'|}}{2\beta_z},
\end{eqnarray}
\end{subequations}
in which $\beta_z = k_0n_{0}(z)$ represents the wavenumber of radiative waves in the $z$ direction. In terms of $G(z,z')$, the radiative waves can then be expressed as
\begin{subequations}
\label{Eq16}
\begin{eqnarray}
\label{Eq16.1}
\Delta E_x(z)=k_0^2(\xi_{0,-1}R_y + \xi_{0,1}S_y) \int_{PC} G(z,z')\Theta_0(z')dz', \nonumber \\
\\
\label{Eq16.2}
\Delta E_y(z)=k_0^2(\xi_{-1,0}R_x + \xi_{1,0}S_x) \int_{PC} G(z,z')\Theta_0(z')dz'. \nonumber \\
\end{eqnarray}
\end{subequations}

By multiplying both sides of the above equations by $\Theta_0^*(z)$ and integrating over the PC region along the $z$ axis, we obtain
\begin{subequations}
\begin{eqnarray}
\label{Eq17.1}
\int_{PC}\Delta E_x(z)\Theta_0^*(z)dz &&= k_0^2(\xi_{0,-1}R_y + \xi_{0,1}S_y)\cdot \\
&&\iint_{PC} G(z,z')\Theta_0(z')\Theta_0^*(z) dz'dz, \nonumber\\
\label{Eq17.2}
\int_{PC}\Delta E_y(z)\Theta_0^*(z)dz &&= k_0^2\cdot(\xi_{-1,0}R_x + \xi_{1,0}S_x)\cdot \\
&&\iint_{PC} G(z,z')\Theta_0(z')\Theta_0^*(z) dz'dz.\nonumber
\end{eqnarray}
\end{subequations}
As a consequence, the second terms of the right-hand sides of the coupled-wave equations (\ref{Eq10.1})-(\ref{Eq10.4}) can be replaced by terms only associated with the basic waves.

Next, we obtain solutions for high-order waves, $E_{x,m,n}(z)$ and $E_{y,m,n}(z)$, where $\sqrt{m^2+n^2}>1$. It is difficult to solve Eqs. (\ref{Eq13.1})-(\ref{Eq13.3}) directly, thus we introduce a proper linear combination of $E_{x,m,n}(z)$ and $E_{y,m,n}(z)$ in order to modify Eqs. (\ref{Eq13.1})-(\ref{Eq13.3}) and obtain a set of equations of the form
\begin{subequations}
\begin{eqnarray}
\label{Eq18.1}
&&[\frac{\partial^2}{\partial
z^2}+k_0^2n_0^2](mE_{x,m,n}+nE_{y,m,n} )=\nonumber\\
&&-k_0^2\sum_{\substack {m'\neq m, \\n' \neq n}} \xi_{\substack {m-m',\\n-n'}}(mE_{x,m',n'}+nE_{y,m',n'}),\\
\label{Eq18.2}
&&[\frac{\partial^2}{\partial
z^2}+k_0^2n_0^2-(m^2+n^2)\beta_0^2](nE_{x,m,n}-mE_{y,m,n} )=\nonumber\\
&&-k_0^2\sum_{\substack {m'\neq m, \\n' \neq n}} \xi_{\substack {m-m',\\n-n'}}(nE_{x,m',n'}-mE_{y,m',n'}),\\
\label{Eq18.3}
&&\frac{\partial}{\partial z}[mE_{x,m,n}+nE_{y,m,n}]=0.
\end{eqnarray}
\end{subequations}
The substitution of Eq. (\ref{Eq18.3}) into Eq. (\ref{Eq18.1}) yields
\begin{eqnarray}
\label{Eq19}
&&n_0^2(mE_{x,m,n}+nE_{y,m,n} )= \nonumber \\
&&-\sum_{\substack {m'\neq m, \\n' \neq n}} \xi_{\substack {m-m',\\n-n'}}(mE_{x,m',n'}+nE_{y,m',n'}).
\end{eqnarray}
It is worthwhile noting that Eq. (\ref{Eq19}) is equivalent to the transversality constraint; i.e., $\nabla\cdot(\bm D(\bm r))=\nabla\cdot(\varepsilon(\bm r)\bm E(\bm r))=0$ must be satisfied.
Next, we solve Eqs. (\ref{Eq18.2}) and (\ref{Eq19}) to obtain solutions for the high-order waves $E_{x,m,n}(z)$ and $E_{y,m,n}(z)$. The linear combination $(nE_{x,m,n}-mE_{y,m,n})$ can be solved from Eq. (\ref{Eq18.2}) by using a similar Green function approach. The Green function $G_{m,n}(z,z')$ satisfies
\begin{eqnarray}
\label{Eq20.1} [\frac{\partial^2}{\partial
z^2}+k_0^2n_0^2-(m^2+n^2)\beta_0^2]G_{m,n}(z,z')=- \delta(z,z'),\nonumber \\
\end{eqnarray}
where
\begin{eqnarray}
\label{Eq20.2}
G_{m,n}(z,z')&&\simeq \frac{e^{-\beta_{z,m,n}|z-z'|}}{2\beta_{z,m,n}}, \nonumber\\ \beta_{z,m,n}&&=\sqrt{(m^2+n^2)\beta_0^2-k_0^2n_0^2(z)}.
\end{eqnarray}
Thus, we obtain
\begin{eqnarray}
\label{Eq21}
(n&&E_{x,m,n}-mE_{y,m,n} )=
k_0^2\sum_{\substack {m'\neq m, \\n' \neq n}} \xi_{\substack {m-m',\\n-n'}} \cdot \nonumber\\
&&\int_{PC}(nE_{x,m',n'}(z')-mE_{y,m',n'}(z'))G_{m,n}(z,z')dz'.\nonumber \\
\end{eqnarray}

\begin{figure*}[htbp]
  \includegraphics[width=10cm]{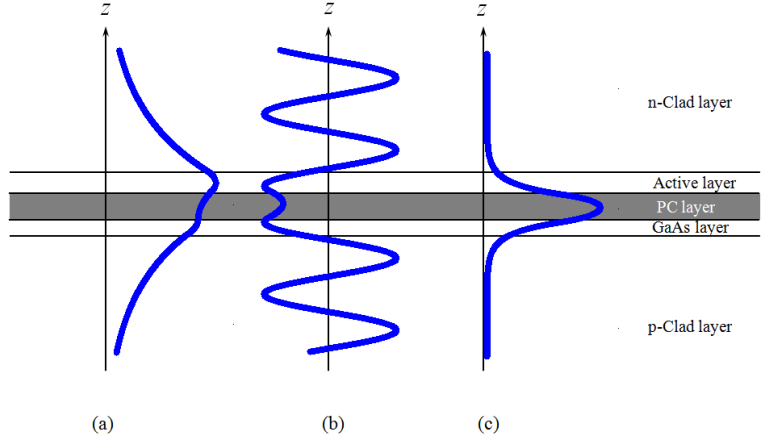}\\
  \caption{(Color online) Schematic illustration of electric-field ($E_x$ or $E_y$) profiles in the vertical direction for wavevectors with different in-plane wavenumbers: (a) basic waves, (b) radiative waves, (c) high-order waves with (m,n)=(1,1). }\label{fig.field-profile}
\end{figure*}

Under the assumption that only the basic waves are important in generating high-order waves, Eqs. (\ref{Eq19}) and (\ref{Eq21}) can be rewritten as
\begin{widetext}
\begin{subequations}
\begin{eqnarray}
\label{Eq22.1}
  mE_{x,m,n}+nE_{y,m,n} &\simeq& -\frac{1}{n_0^2}(n\xi_{\substack{m-1,\\n}}R_x + n\xi_{\substack{m+1,\\n}}S_x + m\xi_{\substack{m,\\n-1}}R_y
  +m\xi_{\substack{m,\\n+1}}S_y)\Theta_0(z) \triangleq E^+(z),\\
\label{Eq22.2}
  nE_{x,m,n}-mE_{y,m,n} &\simeq& k_0^2(-m\xi_{\substack{m-1,\\n}}R_x -m\xi_{\substack{m+1,\\n}}S_x + n\xi_{\substack{m,\\n-1}}R_y
  +n\xi_{\substack{m,\\n+1}}S_y)\int_{PC}G_{m,n}(z,z')\Theta_0(z')dz' \triangleq E^-(z). \nonumber \\
\end{eqnarray}
\end{subequations}
\end{widetext}
We then obtain
\begin{subequations}
\label{Eq23}
\begin{eqnarray}
\label{Eq23.1}
E_{x,m,n}(z) = \frac{mE^+ +
nE^-}{m^2+n^2},\\
\label{Eq23.2}
E_{y,m,n}(z) = \frac{nE^+-
mE^-}{m^2+n^2}.
\end{eqnarray}
\end{subequations}
By multiplying both sides of Eqs. (\ref{Eq23.1})-(\ref{Eq23.2}) by $\Theta_0^*(z)$ and integrating over the PC region, we obtain
\begin{subequations}
\begin{eqnarray}
\label{Eq24.1}
\int_{PC}E_{x,m,n}(z)\Theta_0^*(z)dz = \int_{PC} \frac{mE^+ +
nE^-}{m^2+n^2}\Theta_0^*(z)dz, \nonumber \\
\\
\label{Eq24.2}
\int_{PC}E_{y,m,n}(z)\Theta_0^*(z)dz = \int_{PC}\frac{nE^+-
mE^-}{m^2+n^2}\Theta_0^*(z)dz. \nonumber \\
\end{eqnarray}
\end{subequations}
These equations can be expressed in terms of basic waves (note the definitions of $E^+$ and $E^-$ in Eqs. (\ref{Eq22.1})-(\ref{Eq22.2})).

Finally, the substitution of Eqs. (\ref{Eq17.1})-(\ref{Eq17.2}) and Eqs. (\ref{Eq24.1})-(\ref{Eq24.2}) into the coupled-wave equations (\ref{Eq10.1})-(\ref{Eq10.4}) leads to an eigensystem that can be written in matrix form as
\begin{eqnarray}
\label{Eq25}
(\delta+i\alpha)\left(
                  \begin{array}{c}
                    R_x \\
                    S_x \\
                    R_y \\
                    S_y \\
                  \end{array}
                \right)=\left(
                          \begin{array}{cccc}
                    C_{11} & C_{12} & C_{13} & C_{14} \\
                    C_{21} & C_{22} & C_{23} & C_{24} \\
                    C_{31} & C_{32} & C_{33} & C_{34} \\
                    C_{41} & C_{42} & C_{43} & C_{44} \\
                          \end{array}
                        \right)\left(
                  \begin{array}{c}
                    R_x \\
                    S_x \\
                    R_y \\
                    S_y \\
                  \end{array}
                \right),\nonumber\\
\end{eqnarray}
where the matrix elements $C_{mn}$ with $ (m,n) \in \{1, 2, 3, 4\}$ are dependent on the PC geometry and the multilayer waveguide structure and can be determined analytically.

It is informative to examine the physical interpretation of the coupled-wave equations (\ref{Eq10.1})-(\ref{Eq10.4}).
The first terms on the right-hand sides represent the conventional 1D coupling effects, i.e., the coupling between two counter-propagating basic waves.
The second terms represent coupling between the radiative waves and the basic waves.
Analytical expressions that describe these coupling effects can be obtained by substituting Eqs. (\ref{Eq17.1})-(\ref{Eq17.2}) into the coupled-wave equations (\ref{Eq10.1})-(\ref{Eq10.4}).
For example, by substituting Eq. (\ref{Eq17.2}) into the second term of the right-hand side of Eq. (\ref{Eq10.1}), we obtain
\begin{eqnarray}
\label{Eq26}
-\frac{k_0^2}{2\beta_0P}\xi_{10}\int_{PC}\Delta E_y(z)\Theta_0^*(z)dz=\zeta_{(1,1)}R_x+\zeta_{(1,2)}S_x, \nonumber\\
\end{eqnarray}
where
\begin{subequations}
\begin{eqnarray}
\label{Eq27.1}
\zeta_{(1,1)}&=&-\frac{k_0^4}{2\beta_0P}\xi_{10}\xi_{-10}\iint_{PC}
G(z,z')\Theta_0(z')\Theta_0^*(z)dz'dz, \nonumber\\
\\
\zeta_{(1,2)}&=&-\frac{k_0^4}{2\beta_0P}\xi_{10}\xi_{10}\iint_{PC}
G(z,z')\Theta_0(z')\Theta_0^*(z)dz'dz. \nonumber \\
\end{eqnarray}
\end{subequations}
These are in general complex numbers and are similar in form to the radiation coupling coefficients derived for second-order DFB lasers using conventional CWT \cite{kazarinov}.
The third terms on the right-hand sides of the coupled-wave equations represent the 2D optical coupling of high-order waves.
In the case of TE polarization, basic waves propagating in directions perpendicular to each other within the plane do not couple directly. Coupling instead takes place via high-order waves that propagate at oblique angles, a phenomenon that has been demonstrated by the previous CWT model \cite{sakai1}.
However, the contribution of only four high-order waves was previously taken into account.
The 2D coupling described by our extended coupled-wave equations is far richer in nature;
the infinite summations allow us to include an infinite number of high-order waves, thus making it possible to capture all the important 2D optical coupling effects.
The generalized formalism above is capable of treating air holes of any arbitrary shape by inclusion of the appropriate high-order Fourier components.

Another important feature of our CWT model is that the formulation has been rigorously derived for a 3D structure. Unlike 2D PCs, the structure in this case is \emph{not} uniform in the vertical direction and thus the TE -field profile might be different for each individual wavevector.
As mentioned above, these wavevectors are classified into three groups: basic waves, radiative waves and high-order waves.
The TE -field profiles for these wavevectors are depicted schematically in Fig. \ref{fig.field-profile}.
The profiles were calculated for the waveguide structure shown in Fig. \ref{fig.waveguide} with FF=0.16, by employing Eqs. (\ref{Eq6}), (\ref{Eq16}) and (\ref{Eq23}), respectively (the real part of the electric field is plotted).
It is clearly apparent that for a 3D structure, the individual wavevectors do \emph{not} have identical field profiles.
The basic waves have the same field profile as the fundamental waveguide mode, the amplitude of which has a peak at the active layer and decays slowly towards the upper and lower cladding layers.
The radiative waves possess an oscillating field profile along the $z$ direction and emanate in the direction normal to the PC plane to constitute the laser output (surface emission).
The field profile of the high-order waves is more complicated because it is determined by both Eqs. (\ref{Eq22.1}) and (\ref{Eq22.2}).
We calculated field profiles for several cases with different $(m,n)$ and found that in general, the profile is characterized by Eq. (\ref{Eq22.2}).
A typical profile is shown in Fig. \ref{fig.field-profile}(c) for high-order waves with
$(m,n)=(1,1)$.
It is apparent that the high-order waves are more strongly confined within the PC layer compared to the basic waves, and that they decay evanescently outside the PC layer. This evanescent character is described by the Green function $G_{m,n}(z,z')$ in Eq. (\ref{Eq20.2}) (note that $\beta_m^2+\beta_n^2=(m^2+n^2)\beta_0^2>k_0^2n_0^2(z)$).
In addition, it can easily be found from Eq. (\ref{Eq20.2}) that the field profile is largely dependent on the order of the waves (m,n): the higher the wavevector order, the more strongly the field is confined in the PC layer.
In short, the field profiles for the individual wavevectors in a 3D structure are extremely complicated, which represents the fundamental difference between 3D and 2D systems. Therefore, each field profile must be treated very carefully in order to accurately quantify the coupling effects in a 3D system.
In the previously reported CWT analyses \cite{imada2} \cite{sakai1}, an approximation based on the effective refractive index was used in the 2D calculations.
This approximation implies that all the individual wavevectors have the same field profile in the vertical direction as that of the fundamental waveguide mode, an assumption that might lead to inaccurate results. In the following section, we present detailed numerical examples in order to illustrate this fundamental difference.

\section{Numerical results and discussion}
As described above, we have developed a CWT model that incorporates high-order coupling effects and allows a more accurate definition of the coupling coefficients in a 3D system to be derived.
When Eqs. (\ref{Eq10.1})-(\ref{Eq10.4}) are solved as an eigenvalue problem, we can directly evaluate two most important properties of the band-edge modes for square-lattice PC-SELs with infinite periodic structures, i.e., the mode loss $\alpha$ and the mode frequency $\omega$. In order to understand the effects of asymmetric air-hole shapes, we present numerical results for the three shapes shown in Fig. \ref{fig.sq-lattice}(b): circular (CC),  equilateral triangular (ET), and right-angled isosceles triangular (RIT) shapes.
In all the following calculations, we use a waveguide structure characterized by the parameters in Table \ref{Tab.1}, with the dielectric constants $\varepsilon_a=1.0$, $\varepsilon_b=12.7449$, and the lattice constant $a=295$ nm.

For a square-lattice PC, there are four band-edge modes at the second order $\Gamma$-point for TE polarization \cite{sakai0}. We refer to these as modes A, B, C and D, in ascending order of frequency.
Modes C and D correspond to the symmetric mode of 1D-DFB lasers, and thus have significant loss \cite{miyai2}.
In contrast, modes A and B correspond to the anti-symmetric mode of 1D-DFB lasers, and lase more easily than modes C and D. Therefore, we restrict the following discussion to modes A and B only.
\begin{figure}[hbtp]
\centerline{
\includegraphics[width=7cm]{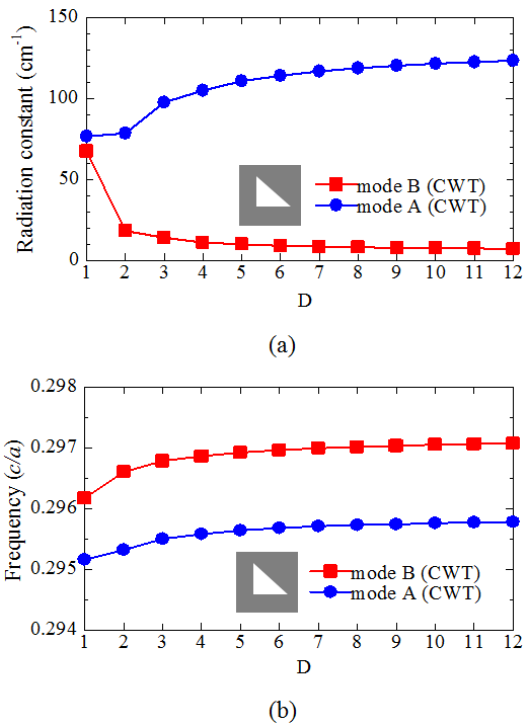}}
 \caption{(Color online) Radiation constant and mode frequency as a function of the truncation order $D$. These plots were calculated for the asymmetric RIT air-hole shape with FF=0.16.}\label{fig.DD}
\end{figure}
In order to solve Eqs. (\ref{Eq10.1})-(\ref{Eq10.4}), we first need to truncate the summations at a certain order of $m$ and $n$. Accordingly, we define a quantity $D$ such that $|m|\leq D, |n|\leq D$.
The effects of truncating the summations can be observed by plotting the radiation constant (a parameter defined by $\alpha_r=2\alpha$ to quantify the modal power loss) and the mode frequency as a function of $D$, as shown in Fig. \ref{fig.DD}. For illustration, we only show results for the asymmetric RIT air-hole shape (FF=0.16), the modeling of which requires the inclusion of many more high-order waves than the CC and ET shapes.
It is apparent from Fig. \ref{fig.DD} that the radiation constant is more sensitive to $D$ than the mode frequency, which indicates that a large number of high-order wavevectors must be included in order to calculate the radiation constant accurately.
Both the radiation constant and mode frequency change little for $D \geq 10$, hence we use a truncation of $D=10$ in all of the following calculations.

\begin{figure*}[hbtp]
\centerline{
\includegraphics[width=18cm]{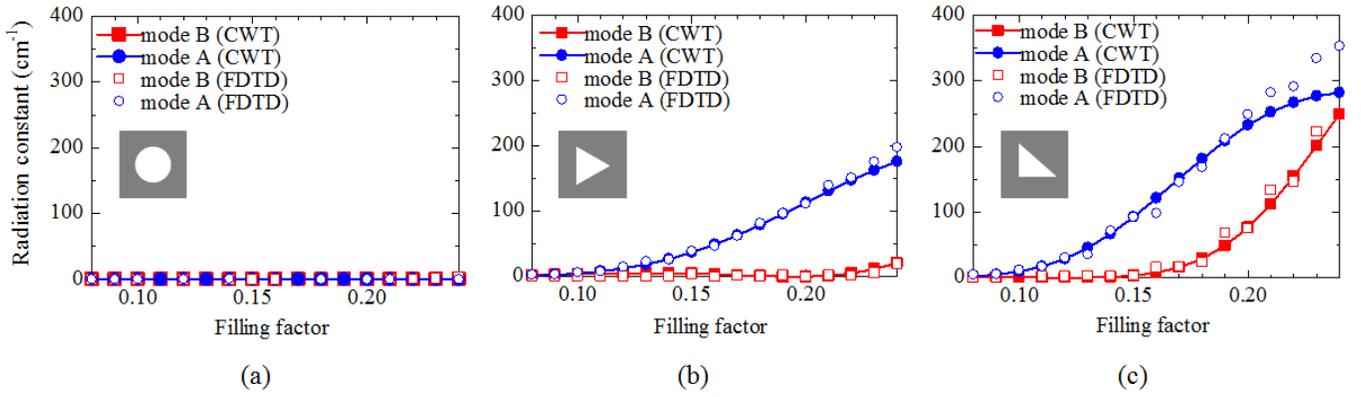}}
 \caption{(Color online) Radiation constant as a function of FF for (a) CC, (b) ET, and (c) RIT air-hole shapes. Some 3D-FDTD points are missing in the case of the CC shape because $Q$ is infinitely large.}\label{fig.rad}
\end{figure*}
\begin{figure*}[hbtp]
\centerline{
\includegraphics[width=18cm]{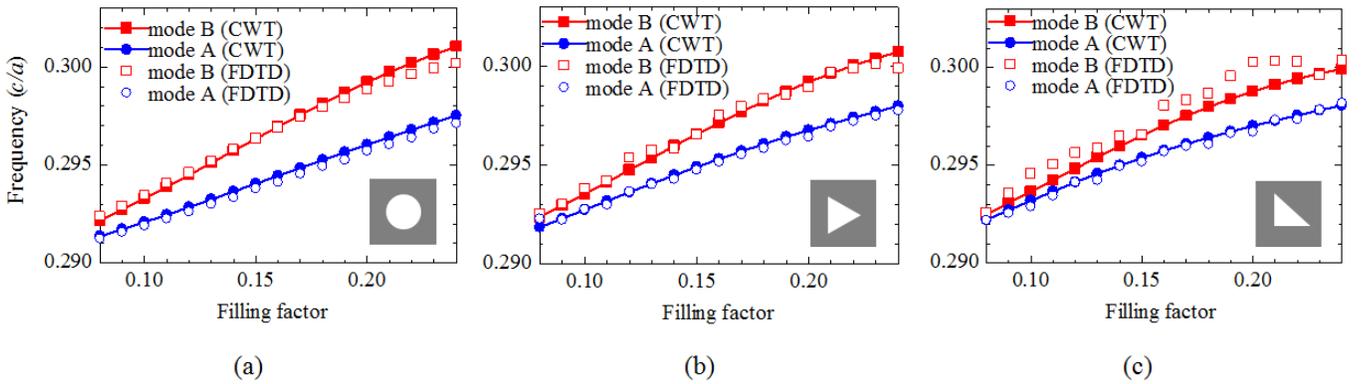}}
 \caption{(Color online) Mode frequency as a function of FF for (a) CC, (b) ET, and (c) RIT air-hole shapes.}\label{fig.fre}
\end{figure*}

In order to confirm the accuracy of the above CWT analysis results, we also performed 3D-FDTD simulations \cite{yokoyama,oskooi} for the structure shown in Fig. \ref{fig.waveguide}.
We used a computational cell of $40\times40\times640$ pixels ($x\times y \times z$), corresponding to $1\times1\times16$ lattice periods, with absorbing boundary layers in the $z$ direction and
periodic boundary conditions in $x$ and $y$.
The $Q$ factor obtained by the 3D-FDTD method was used to compute the radiation constant via the following relationship \cite{rosenblatt}:
\begin{eqnarray}
\label{Eq28}
\alpha_r\simeq\frac{\beta_0}{Q}=\frac{2\pi/a}{Q}.
\end{eqnarray}
Figures \ref{fig.rad} and \ref{fig.fre} show the radiation constant and mode frequency as a function of FF, obtained by both the CWT and 3D-FDTD methods.
It is clear that the CWT results are in good agreement with the 3D-FDTD simulations. The slight deviation of the 3D-FDTD data in the case of the RIT air hole can be attributed mainly to numerical effects (i.e., low resolution).
However, the calculation time required for the two methods is markedly different.
For a specific FF, the 3D-FDTD simulation takes $\sim 4$ hours using a supercomputer system
(64 cores and 9.0 GB memory), whereas the CWT analysis takes less than $1$ second with
a personal computer (1 core@2.20GHz and negligible memory usage).
Although a large number of wavevectors were included in the CWT analysis,
the calculation time is nevertheless short due to the semi-analytical nature of the algorithm.

\begin{figure}[hbtp]
\centerline{
\includegraphics[width=9cm]{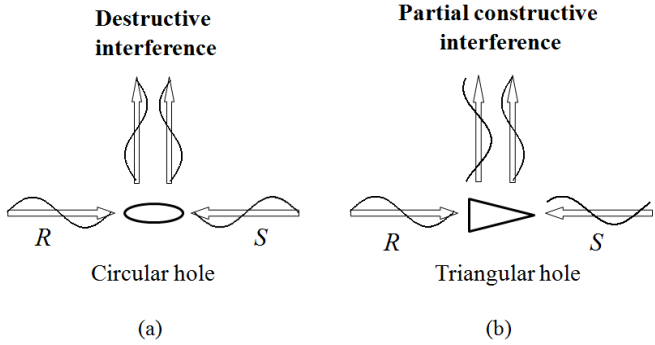}}
 \caption{Schematic illustration of the interference occurring in the vertical direction for (a) circular holes and (b) triangular holes. The arrows represent the propagation directions of the basic waves and radiative waves, and the sine curves represent the phases of the fields for these waves.}\label{fig.interf}
\end{figure}

It is noteworthy that the radiation constant for the symmetric CC air-hole shape is zero (corresponding to an infinite $Q$ factor in the 3D-FDTD method) at every value of FF, whereas it increases with FF when the air-hole shapes are asymmetric (ET or RIT). This difference can be physically interpreted by considering the interference occurring in the vertical direction, as illustrated in Fig. \ref{fig.interf}.
The radiation field depends on the phase difference of the waves diffracted vertically, which arise from counter-propagating basic waves in the PC plane.
In the case of the symmetric CC air-hole shape, the two basic waves propagating in the $x$ or $y$ direction are intrinsically out of phase \cite{miyai2} and their diffracted waves thus have a phase difference of $\pi$. Therefore, destructive interference occurs and the two diffracted waves cancel each other out.
In contrast, when the air-hole shape is asymmetric, such as for the ET and RIT shapes,
the counter-propagating basic waves are no longer out of phase and the phase difference of the diffracted waves will deviate from $\pi$. The destructive interference is consequently suppressed, giving rise to partial constructive interference.
Therefore, a higher output power can be expected when asymmetric air-hole shapes are used.
\begin{figure}[hbtp]
\centerline{
\includegraphics[width=7cm]{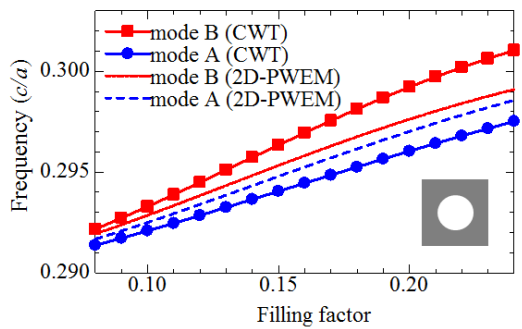}}
 \caption{(Color online) Mode frequency for CC air holes as a function of FF, calculated using the CWT and 2D-PWEM approaches.}\label{fig.pwem}
\end{figure}

We note that the mode frequency plotted in Fig. \ref{fig.fre} was calculated for a 3D structure, which should be different from that calculated for a 2D structure.
In order to elucidate the difference between 3D and 2D calculations, we evaluated the mode frequency for the CC air-hole shape using the 2D-PWEM approach \cite{imada2}, which is plotted versus FF in Fig. \ref{fig.pwem} together with the CWT results of Fig. \ref{fig.fre}(a).
A total of 225 plane waves were used in the 2D-PWEM calculation and the effective refractive index approximation \cite{sakai0} was employed.
Comparing the results for the two methods, it is apparent that the mode gap between modes A and B ($\omega_B-\omega_A$) calculated using the present CWT model is significantly larger than that calculated using the 2D-PWEM approach.
We suggest that a physical explanation can be found by considering the fact that in a 3D structure, the high-order waves (generated by basic waves) are more strongly confined in the PC layer, as depicted in Fig. \ref{fig.field-profile}(c).
The effective refractive index approximation used in the 2D-PWEM calculation implicitly assumes that the transverse field profile of all the high-order waves is identical to that of the fundamental waveguide mode, the amplitude of which slowly decays outside the PC layer, as shown in Fig. \ref{fig.field-profile}(a).
Therefore, the 2D optical coupling strength is greater in a 3D structure than in a 2D structure, leading to a larger mode gap in a 3D structure.

\section{Summary}
In summary, we have presented a generalized CWT model for 2D square-lattice PC-SELs with TE polarization. Our model incorporates a large number of high-order wavevectors, and we have derived a general and rigorous formulation to describe the coupling effects that occur in a 3D system. Moreover, our general coupled-wave formulation can be applied to air holes of arbitrary shape.
 The accuracy of our CWT model has been confirmed by comparison with 3D-FDTD simulations, which require significantly greater computational time.

In our CWT analysis, we have shown that the inclusion of a sufficiently large number of high-order wavevectors is important for an accurate study of the band-edge modes, especially in the case of asymmetric air-hole shapes.
Furthermore, a 3D characterization allows realistic laser devices to be more accurately modeled than in a 2D analysis.
We have also discussed the fundamental differences between 3D and 2D systems.
By comparison with the results obtained for a 2D system using the 2D-PWEM approach, a larger mode gap was calculated for the 3D structure modeled using the CWT method.
This larger mode gap can mainly be attributed to a much stronger 2D optical coupling via high-order wavevectors, which are found to be strongly confined in the PC layer.
By evaluating the radiation constant and mode frequency of the band-edge modes for several different air-hole shapes, we have found that asymmetric air holes are beneficial for improving the output power of 2D PC-SELs because destructive interference in the vertical direction can be suppressed.

As the purpose of this work is to develop a CWT model that better describes the coupling effects in 2D square-lattice PC lasers with TE polarization, we have restricted our analysis to infinite periodic structures.
However, the theoretical framework constructed here can be extended to the analysis of finite structures without modifying the basic methodology. Moreover, extension of our theory to TM polarization, triangular-lattice PCs, and more complicated PC geometries (such as PCs comprised of air-holes with tilted sidewalls) is straightforward.
We believe that further theoretical work based on this framework will enable efficient optimization of the
structures of 2D PC-SELs for a range of applications, as well as providing a powerful analytical tool to comprehensively understand the properties of 2D PC-SELs.

\section*{Acknowledgments}
The authors are grateful to Dr. M. Yamaguchi, Dr. A. Oskooi, Dr. J. Upham, and Dr. B. S. Song for helpful discussions and valuable suggestions. This work was partly supported by the Core Research for Evolution Science and Technology program of the Japan Science and Technology Agency (CREST-JST) and by the Global COE Program. Three of the authors (Y. Liang, C. Peng and S. Iwahashi) are supported by Research Fellowships of the Japan Society for Promotion of Science.

\end{document}